# Charge detection of a quantum dot under different tunneling barrier symmetries and bias voltages

Jingwei Mu[1,3], Weijie Li[1,3], Shaoyun Huang[1], Dong Pan[2,4], Yuanjie Chen[1], Ji-Yin Wang[1], Jianhua Zhao[2,4], and H. Q. Xu[1,3,4,*]

[1]*Beijing Key Laboratory of Quantum Devices, Key Laboratory for the Physics and Chemistry of Nanodevices and Department of Electronics, Peking University, Beijing 100871, China*

[2]*State Key Laboratory of Superlattices and Microstructures, Institute of Semiconductors, Chinese Academy of Sciences, P.O. Box 912, Beijing 100083, China*

[3]*Academy for Advanced Interdisciplinary Studies, Peking University, Beijing 100871, China*

[4]*Beijing Academy of Quantum Information Sciences, Beijing 100193, China*

[*]To whom correspondence should be addressed. Email: hqxu@pku.edu.cn

(May 22, 2020)

## Abstract

We report on the realization of a coupled quantum dot (QD) system containing two single QDs made in two adjacent InAs nanowires. One QD (sensor QD) is used as a charge sensor to detect the charge state transition in the other QD (target QD). We investigate the effect of the tunneling barrier asymmetry of the target QD on the detection visibility of charge state transition in the target QD. The charge stability diagrams of the target QD under different configurations of barrier-gate voltages are simultaneously measured via the direct signals of electron transport through the target QD and via the detection signals of charge state transition in the target QD from the sensor QD. We find that the complete Coulomb diamond boundaries of the target QD and the transport processes involving the excited states in the target QD can be observed in the transconductance signals of the sensor QD only when the tunneling barriers of the target QD are nearly symmetric. These phenomena are explained by analyzing the effect of the ratio of the two tunneling rates on the electron transport processes through the target QD. Our results imply that it is important to consider the symmetry of the tunneling couplings when constructing a charge sensor integrated QD device or qubit.

# I. Introduction

Semiconductor quantum dots (QDs) are among the most promising systems for applications in solid state-based quantum computing [1-3] and quantum simulation [4,5]. Especially, QDs made in narrow bandgap semiconductor nanowires, such as InAs and InSb nanowires, possess large Landé g-factors [6-9] and strong spin-orbit interaction [8,10-13], which can be employed for rapid manipulations of spin states by all-electrical means [14-17]. Coherent manipulations of single spin qubits have been demonstrated by direct current measurements of InAs nanowire QDs [15,18]. Further studies of real-time readout of spin qubits via a spin-to-charge conversion and of the physics of a QD in the few-electron regime [19,20] require integrations of highly sensitive charge detectors [21-30]. Thus, realization of sensitive charge detector integrated QDs is a crucial step toward integration of spin qubits [31,32] for quantum computation and quantum simulation [4].

Previous works have demonstrated charge detection of QDs in the linear response regime [23-28]. Charge detection of QDs at finite source-drain bias voltages is a more complicated process due to involving excited states. Although some excited states in a QD have been detected via an integrated charge sensor in some specific conditions [33-36], detection of a complete excited state spectrum in a QD via a charge sensor has rarely been reported. To understand this problem, since charge states in a QD at a finite bias voltage largely depend on the symmetry of the tunneling couplings of QD and on the polarity of the bias voltage, it is necessary to perform full experimental measurements and analyses for the effect of the symmetry of the tunneling couplings on the detection of the charge states in the QD at finite source-drain bias voltages.

In this article, we demonstrate that an excited state spectrum of a QD can only be observed when the tunneling couplings of the QD states to the source and drain electrodes are sufficiently symmetric. We employ a system of a single QD (target QD) integrated with a single QD charge detector (sensor QD). The device is realized in two adjacent InAs nanowires via fine finger gate technique. Here, InAs nanowires are used, because it has been shown that nanowire QDs defined by fine finger gates have advantages in tunability of individual tunneling barriers [37-40]. The capacitive coupling between the two QDs is enhanced through a thin metal gate. Thus, the sensor QD is highly sensitive to detect the charge state transitions of the target QD even in the cases where the direct current signal of the target QD is too weak to be detectable. The charge stability diagrams of the target QD are measured simultaneously via the direct current signals of the target QD and via the detection signals of the sensor QD. We find that the simultaneous signals of the sensor QD could not map out the complete Coulomb diamond boundaries of the target QD when the tunneling couplings of the target QD are distinctly asymmetric. In addition, we find that the signatures of the excited states in the target QD can be observed in the detection signals of the sensor QD only when the tunneling couplings of the target QD are nearly symmetric. We analyze in detail the electron transport processes in the target QD under different tunneling coupling configurations and bias voltage polarities, and discuss the physical origin for the experimental observations.

**II. Methods**

The InAs nanowires employed in this work are grown by molecular beam epitaxy (MBE) and have a diameter of ~30 nm [41]. For device fabrication, the MBE-grown InAs nanowires are transferred by a dry method from the growth substrate onto a heavily p-doped Si substrate capped with a 300-nm-thick layer of $SiO_2$. The Si substrate and the $SiO_2$ layer serve as a global back gate and a gate dielectric layer, respectively. Here, pairs of adjacent InAs nanowires are selected for further device fabrication. The source and drain contact areas of the nanowires are defined through electron-beam lithography (EBL) and are chemically etched in a diluted $(NH_4)_2S_x$ solution to remove the oxides on the exposed surfaces of the nanowires. Subsequently, the source and drain electrodes are fabricated by deposition of 5-nm-thick titanium and 90-nm-thick gold via electron-beam evaporation (EBE) and lift-off. A 10-nm-thick $HfO_2$ layer is then deposited onto the sample via atomic layer deposition (ALD). Finally, top finger gates together with coupling metal gates with an averaged width of ~30 nm are fabricated by a combined step of EBL, EBE of 5-nm-thick titanium and 25-nm-thick gold, and lift-off. Figure 1(a) shows a scanning electron microscope (SEM) image of a fabricated device investigated in this work. The electrode array on top of the left InAs nanowire consists of eleven finger gates with a pitch of ~60 nm, while the array on top of the right InAs nanowire consists of three finger gates with a pitch of ~70 nm.

The cryogenic transport measurements are performed in a $He^3/He^4$ dilution refrigerator at a base temperature of ~20 mK. The global back-gate voltage is set at $V_{bg}$=8 V in order to set both the two InAs nanowires at an n-type conduction state throughout the measurements. The transfer characteristics of the finger gates are measured and a current pinch-off voltage for each individual gate is found to be in a range of −0.85 to −0.4 V. A finite tunneling barrier can be formed by applying a negative voltage to each of these finger gates around its pinch-off voltage. In this way, a single QD can be defined using two finger gates to form two tunneling barriers in an InAs nanowire.

### III. Results and discussion

*A. Capacitive coupling between the sensor QD and the target QD*

Figure 1(b) displays the differential conductance $dI_{DS}/dV_{DS}$ of a single QD (target QD) defined in the right InAs nanowire shown in Fig. 1(a) as a function of source-drain bias voltage $V_{DS}$ and voltage $V_{G2}$ applied to finger gate G2 (charge stability diagram). Here, gates G1 and G2 are used to form the two tunneling barriers of the target QD. Gate G2 is also used to tune the electrostatic potential of the target QD. The regular Coulomb diamonds as well as the close points seen at zero $V_{DS}$ between neighboring Coulomb diamonds indicate the formation of a well-defined single QD. An averaged electron addition energy of ~5 meV can be extracted from the measured charge stability diagram. Figure 1(c) shows the differential conductance $dI_{ds}/dV_{ds}$ of a single QD (sensor QD) defined in the left InAs nanowire shown in Fig. 1(a) as a function of source-drain bias voltage $V_{ds}$ and voltage $V_{g2}$ applied to finger gate g2. Gates g1 and g3 are used to form the two tunneling barriers of the sensor QD. Gate g2 is employed to tune the electron number in the sensor QD. An averaged electron addition energy of ~3 meV can be extracted from the charge stability diagram.

The sensor QD is adjacent to the target QD and the capacitive coupling between them is enhanced through a thin metal gate M as indicated in Fig. 1(a). Therefore, these two QDs can be regarded as a parallel double QD (DQD) system with a finite inter-dot capacitive coupling and with no tunneling coupling [42]. In order to investigate the capacitive coupling strength between the two QDs, we simultaneously measure the source-drain currents $I_{DS}$ of the target QD and $I_{ds}$ of the sensor QD as a function of voltages $V_{G2}$ and $V_{g2}$ at $V_{DS}$=50 µV and $V_{ds}$=50 µV. The results are shown in Fig. 2(a), in which the current $I = I_{ds} + I_{DS}$ is plotted. Here, two current lines of the target QD and three current lines of the sensor QD are seen. These current lines exhibit finite slopes, which mean that both gates G2 and g2 capacitively couple to the target QD and the sensor QD. We also observe anti-crosses in the regions where two current lines belonging to different QDs intersect. Figure 2(b) shows a plot of the current $I_{DS}$ of the target QD only, where the anti-crosses seen in Fig. 2(a) are now seen as line breakings [see the regions marked by the white circles in Fig. 2(b)]. In an electrostatic capacitance network model for a DQD, the strength of an anti-cross represents the strength of the capacitive coupling between the two QDs [43]. Therefore, we can infer that the capacitive coupling between the sensor QD and the target QD is strong enough to enable the detection of the charge state transitions in the target QD.

### B. Detection of charge state transitions in the target QD at a small bias voltage by the sensor QD

Figure 3(a) shows the Coulomb current oscillations of the sensor QD as a function of $V_{g2}$ at $V_{ds}$=0.2 mV. Each current peak corresponds to a change in the charge occupation of the sensor QD by one electron. The current $I_{ds}$ of the sensor QD is expected to be sensitive to the charge state transitions in the target QD by setting $V_{g2}$ at a steep slope of a Coulomb current peak of the sensor QD. Figure 3(b) shows the current $I_{DS}$ of the target QD and the simultaneously measured current $I_{ds}$ of the sensor QD as a function of $V_{G2}$ at $V_{DS}$=0.2 mV and $V_{ds}$=0.2 mV. Here, $V_{g2}$ is fixed at the declining slope of a Coulomb current peak of the sensor QD, as indicated by a green star in Fig. 3(a). The current $I_{ds}$ increases gradually with decreasing $V_{G2}$ due to the capacitive coupling between the sensor QD and gate G2 as discussed above. Besides, distinct jumps in the current $I_{ds}$ of the sensor QD happen at voltages $V_{G2}$ where the current $I_{DS}$ of the target QD is at Coulomb current oscillation peaks. Each electron tunneling out of the target QD can effectively change the electrostatic potential of the sensor QD, causing a distinct decrease in the current $I_{ds}$ with decreasing $V_{G2}$ [23]. Thus, the charge state transitions in the target QD can be well detected by tracking the abrupt changes in the current $I_{ds}$. We also observe that no more Coulomb current peak is visible at voltage $V_{G2}$ lower than −1.125 V, where the direct current signal of the target QD may be too weak to be detectable due to the opaque tunneling barriers of the target QD. Remarkably, the current $I_{ds}$ of the sensor QD still shows distinct jumps, indicating that the sensor QD is highly sensitive and has the potential to assist us to access the few-electron regime or even the zero-electron regime of the target QD.

### C. Detection of ground states of the target QD at a large bias voltage by the sensor QD

Figures 4(a)-4(c) show the charge stability diagrams of the target QD in three successive voltage ranges of $V_{G2}$. Here, the voltage $V_{G1}$ applied to gate G1 is fixed at $V_{G1}=-0.8$ V. Figures 4(d)-4(f) show the simultaneously measured corresponding transconductance signals $dI_{ds}/dV_{G2}$ of the sensor QD, respectively. The high transconductance lines with a positive (negative) slope correspond to the alignments of the electron states in the target QD with the electrochemical potential $\mu_D$ ($\mu_S$) of the drain (source) electrode [43]. It is seen that the direct transport measurements of the target QD show complete Coulomb diamonds in Fig. 4(a), while the simultaneously measured signals of the sensor QD only reveal the high transconductance lines with a positive slope in Fig. 4(d). Assuming that the target QD has been filled with N-1 electrons, we consider two successive ground states G(N-1) and G(N) in the target QD and the process of tuning the two states through the source-drain bias window by gate G2. Here, we only analyze the case of single-level transport, i.e., $e|V_{DS}|<\Delta$, where $\Delta$ is an electron excitation energy in the target QD. The mean dwell time of the Nth electron in the target QD is determined by the ratio of the electron tunneling rates from the source and drain electrodes to the ground state G(N) [33,34]. The tunneling barrier between the target QD and the source electrode is opaquer than the one between the target QD and the drain electrode at voltage $V_{G2}$ lower than −0.78 V, resulting in a much smaller source-QD tunneling rate $\Gamma_S$ than drain-QD tunneling rate $\Gamma_D$ and thus different dwell times of the Nth electron in the target QD at different polarities of the source-drain bias voltage and the energy level configurations of the target QD. Let us first consider a case with $V_{DS}$ set at a positive value as marked by the upper dashed lines in Figs. 4(a) and 4(d). As illustrated in the upper left diagram of Fig. 4(g), when ground state G(N) is in the source-drain bias window, electrons occupying the ground state from the source electrode can immediately transfer to the drain electrode. Thus, the time-averaged electron number

in the target QD remains approximately at N-1 and does not distinctly change when G(N) is shifted across $\mu_S$ with increasing $V_{G2}$. As a consequence, no charge state transition can be clearly detected by the sensor QD as shown in Fig. 4(d), even though it can be clearly observed by the direct electron transport measurements as shown in Fig. 4(a). However, as illustrated in the upper right diagram of Fig. 4(g), when G(N) is shifted below $\mu_D$ by increasing $V_{G2}$, an Nth electron entering ground state G(N) of the target QD will be trapped and the time-averaged electron number in the target QD will change to N. Thus, a distinct change in the number of electrons in the target QD will occur when G(N) is shifted across $\mu_D$, leading to a detectable change in the transconductance of the sensor QD and an overall high transconductance line as shown in Fig. 4(d). Let us now consider a case with $V_{DS}$ set at a negative value as marked by the lower dashed lines in Figs. 4(a) and 4(d). As illustrated in the lower left diagram of Fig. 4(g), when G(N) is in the source-drain bias window, an electron leaving ground state G(N) of the target QD can be immediately replaced by an electron from the drain electrode. Thus, the time-averaged electron number in the target QD is distinctly altered from N-1 to N when G(N) is shifted across $\mu_D$, leading to the observation of a change in the transconductance and an overall high transconductance line with a positive slope in the simultaneously measured detection signals of the sensor QD as seen in Fig. 4(d). As illustrated in the lower right diagram of Fig. 4(g), the effective number of electrons in the target QD will not change and will remain effectively at a value of N when G(N) is shifted across $\mu_S$, leading to no detectable change in the transconductance of the sensor QD as shown in Fig. 4(d). The same analysis can be made for the measurements shown in Figs. 4(c) and 4(f), where the tunneling barrier on the drain side is opaquer than the one on the source side, i.e., $\Gamma_S \gg \Gamma_D$, as illustrated in the schematic energy diagrams shown in Fig. 4(i). However, here the high transconductance lines with a negative slope in the simultaneously measured signals of the sensor QD are observed [see Fig. 4(f)]. For a symmetric case with two nearly equal tunneling barriers of the target QD ($\Gamma_S \approx \Gamma_D$), the time-averaged electron number in the target QD can be effectively altered when G(N) is shifted across $\mu_S$ or $\mu_D$ [see the schematic energy diagrams in Fig. 4(h)]. Therefore, two groups of high transconductance lines with both positive and negative slopes can be observed in the simultaneously measured signals of the sensor QD as seen in Fig. 4(e).

*D. Detection of excited states of the target QD by the sensor QD*

When a high bias voltage $V_{DS}$ (i.e., $e|V_{DS}|\geq\Delta$) is applied, the situation will be more complicated because some excited states in the target QD might contribute to the electron transport and affect the time-averaged electron number in the target QD. We now analyze these situations where the charge state transitions involve the excited states in the target QD. We should already emphasize that the most striking results shown in Figs. 4(d) and 4(f) are that for an asymmetric target QD with two largely different tunneling barriers, no signatures of excited states in the target QD are observed in the detection signals of the sensor QD, although these excited states are clearly observed in the charge stability diagrams of the target QD in the direct transport measurements as shown in Figs. 4(a) and 4(c). To understand these striking results, we draw in Fig. 5(a) a schematic for the charge stability diagram of the target QD around the degenerate N-1 and N electron occupation point [44] and in Figs. 5(b) and 5(c) schematics for the charge state energy levels in the target QD under large positive and negative bias voltages. Here, only two excited states E(N-1) and E(N) are considered for clarity, and the tunneling barrier between the target QD and the source electrode is opaquer than the one between the target QD and the drain electrode. In Fig. 5(a), the blue, red, and orange lines mark the critical points where the charge state transitions of G(N-1)↔G(N), G(N-1)↔E(N), and E(N-1)↔G(N) take place, respectively. Let us first analyze a case with $V_{G2}$ set at a value on the left side of the G(N-1)↔G(N) degenerate point at zero bias voltage as marked by the left vertical gray dashed line in Fig. 5(a). For a positive bias voltage $V_{DS}>0$, as illustrated in the left diagram of Fig. 5(b), when the N electron ground state G(N) and excited state E(N) are in the source-drain bias window, electrons occupying the ground state G(N) and the excited state E(N) from the source electrode can immediately transfer to the drain electrode. Thus, the time-averaged electron number in the target QD remains approximately at N-1. As a consequence, the electron number in the target QD does not have a distinct change when the ground state G(N) and the excited state E(N) are shifted into the source-drain bias voltage window with increasing $V_{G2}$. For a negative bias voltage $V_{DS}<0$, as illustrated in the right diagram of Fig. 5(b), when an Nth electron enters the ground state G(N) from the drain electrode, this electron will be trapped in the target QD for relatively much long time and the excited state E(N) will not be available for another electron to enter. Thus, the time-averaged electron number in the target QD will remain approximately unchanged when the excited state E(N) is moved into the source-drain bias voltage window with increasing $V_{G2}$. Note that a co-tunneling process could occur in which an electron in the ground state G(N) can tunnel to the source electrode when an electron from the drain electrode tunnels into the excited state E(N). However, in this case, the number of electrons in the target QD still remains as N and no change in the electron number in the target QD could be effectively detected by the sensor QD when E(N) is moved into the source-drain bias voltage window with increasing $V_{G2}$. The above analyses explain our observations as shown in Fig. 4(d) that the red lines in Fig. 5(a) are invisible in the transconductance signals of the sensor QD.

Let us now consider a case with $V_{G2}$ set at a value on the right side of the G(N-1)↔G(N) degenerate point at zero bias voltage as marked by the right vertical gray dashed line in Fig. 5(a). For a positive bias voltage $V_{DS}$>0, as illustrated in the left diagram of Fig. 5(c), the excited state E(N-1) is not available for an electron to enter when the ground state G(N-1) is fully occupied. Even in the case when an electron in the ground state G(N-1) is temporarily excited out of the target QD and at the same time an electron enters the excited state E(N-1) for a short time (co-tunneling process), the electron number in the target QD remains unchanged regardless of whether the excited states is moved in or out of the source-drain bias voltage window. For a negative bias voltage $V_{DS}$<0, the same situation will occur, as illustrated in the right diagram of Fig. 5(c), namely that the excited state E(N-1) is, in most time, not available for an electron to enter or could only be occupied when an electron in the ground state G(N-1) is briefly excited out of the target QD. Thus, the electron number in the target QD will not be effectively changed when the excited state E(N-1) is moved in or out of the source-drain bias voltage window. These analyses explain that the orange lines in Fig. 5(a) are difficult to be observed in the transconductance signals of the sensor QD in agreement with our measurements shown in Fig. 4(d). If the tunneling barrier on the drain side is opaquer than the one on the source side, the same analyses can be made and the invisibility of the excited states in the detection signal measurements shown in Fig. 4(f) can be explained. Overall, we arrive at that for an asymmetric target QD with two largely different tunneling barriers, the charge state transitions involving the excited states in the target QD usually could not be observed in the detection measurements via a charge sensor QD.

For a symmetric target QD with two nearly equal tunneling barriers, the above analyses about the visibility of the excited states in the measured detection signals of the sensor QD may not be immediately applicable, due to the fact that the electron tunneling rates through the two tunneling barriers become comparable. As illustrated in Fig. 5(d), when both ground state G(N) and excited state E(N) are in the source-drain bias voltage window, the mean dwell time of the Nth electron in the target QD is no longer zero or sufficiently large. Thus, the time-averaged electron number in the target QD will be different for the cases with the excited state E(N) located inside and outside the source-drain bias voltage window. As a consequence, the excited state E(N) should become observable in the measured charge state detection signals of the sensor QD. In addition, this result should also be independent of whether the source-drain bias voltage is positive or negative as clearly seen in the two schematics of Fig. 5(d). However, we should note that as seen in the left black rectangle in Fig. 4(e), the excited state E(N) is only observed in the detection signal measurements of the sensor QD in the case with negative bias voltages applied to the target QD. At the moment, we do not have a definite explanation for no visibility of the excited state E(N) in our charge state detection signal measurements with a positive bias voltage applied to the target QD. One possible explanation could be due to the presence of a significant difference between the two tunneling rates of an electron to or from the excited state E(N) through the two tunneling barriers because of the complex nature of the wave function of the excited state. Figure 5(e) illustrates the case shown in the right black rectangle of Fig. 4(e), where the detectability of the excited state E(N-1) is in consideration. According to the energy level diagrams shown in Fig. 5(e), an electron can tunnel into the excited state E(N-1) while an electron in the ground state G(N-1) is excited out of the target QD (co-tunneling process). When the excited state E(N-1) is located in the source-drain bias voltage window, the electron in the state will have a possibility of tunneling out the target QD, and thus the dwell time of the electron will be finite, leading to a noticeable change in the time-averaged electron number in the target QD when comparing to the case where E(N-1) is located below the source-drain bias voltage window. As a result, the excited state E(N-1) is observable in the charge state detection signal measurements via the sensor QD, regardless of whether the source-drain bias voltage is positive or negative, which is fully in line with our measurements shown in the right black rectangle of Fig. 4(e).

## IV. Conclusion

In summary, we have performed the measurements of the transport characteristics of an InAs nanowire single QD using a QD charge sensor made in an adjacent InAs nanowire. The two QDs are connected by a thin metal gate, which enhances the capacitive coupling between them. The charge stability diagrams of the target QD are simultaneously measured via the direct current signals of the target QD and via the charge detection signals of the sensor QD. We find that for the target QD with two nearly equal tunneling barriers (a symmetric target QD case), the complete Coulomb diamond boundaries of the target QD and the signatures of the electron transport involving the excited states in the target QD are observable by the charge detection measurements via a charge sensor. However, when the target QD is tuned to be the case with two largely asymmetric tunneling barriers, only part of the Coulomb diamond boundaries is visible in the charge detection signals of the charge sensor, and no signatures of the excited states in the target QD are visible through the charge detector. These striking observations are analyzed by considering the effective electron occupation numbers in the target QD under different tunneling coupling configurations of the target QD and different polarities of the applied source-drain bias voltage. Our study suggests that it is crucial to consider the symmetry of the tunneling couplings when constructing a charge sensor integrated QD for applications in quantum computation and quantum simulation.

**ACKNOWLEDGEMENTS**

This work is supported by the Ministry of Science and Technology of China through the National Key Research and Development Program of China (Grant Nos. 2017YFA0303304, 2016YFA0300601, 2017YFA0204901, and 2016YFA0300802), the National Natural Science Foundation of China (Grant Nos. 91221202, 91421303, 11874071, 11974030, 92065106 and 61974138), the Beijing Academy of Quantum Information Sciences (No. Y18G22), and the Beijing Natural Science Foundation (Grant Nos. 1202010 and 1192017). DP also acknowledges the support from Youth Innovation Promotion Association, Chinese Academy of Sciences (No. 2017156).

CAPTIONS

**FIG. 1.** (a) SEM image (in false color) of a coupled QD system containing a single QD labeled as the target QD and another single QD labeled as the sensor QD made in two adjacent InAs nanowires by fine finger gate technology. Finger gates G1 and G2 are used to define the target QD and finger gates g1 to g3 are employed to define the sensor QD. A thin metal gate labeled as M is fabricated to enhance the capacitive coupling between the two QDs. (b) Charge stability diagram of the target QD. Here, in the measurements, the voltage $V_{G1}$ applied to gate G1 is set at $V_{G1}=-0.8$ V. (c) Charge stability diagram of the sensor QD defined by setting the voltages applied to gates g1 and g3 at $V_{g1}=-0.5$ V and $V_{g3}=-0.45$ V.

**FIG. 2.** (a) Current $I=I_{DS}+I_{ds}$ of the QD charge sensor integrated QD device measured at $V_{DS}=50$ μV and $V_{ds}=50$ μV as a function of voltages $V_{G2}$ and $V_{g2}$ applied to gate G2 and g2. Here, $I_{DS}$ is the current passing through the target QD and $I_{ds}$ is the current passing through the sensor QD. The voltage applied to gate G1 is set at $V_{G1}=-0.8$ V. The voltages applied to gates g1 and g3 are set at $V_{g1}=-0.49$ V and $V_{g3}=-0.45$ V. (b) The same as (a) but only the current $I_{DS}$ of the target QD is plotted. White circles indicate the anti-crossing regions of the current lines.

**FIG. 3.** (a) Current $I_{ds}$ of the sensor QD measured at $V_{ds}=0.2$ mV as a function of voltage $V_{g2}$ applied to gate g2. The sensor QD is defined by setting $V_{g1}=-0.49$ V and $V_{g3}=-0.45$ V. The green star indicates the position at which subsequent charge detection measurements are performed. (b) Current $I_{DS}$ of the target QD and simultaneously measured current $I_{ds}$ of the sensor QD at $V_{DS}=0.2$ mV and $V_{ds}=0.2$ mV as a function of voltage $V_{G2}$ applied to gate G2. The voltage applied to gate G1 is set at $V_{G1}=-0.8$ V. Distinct jumps in current $I_{ds}$ appear at voltages $V_{G2}$ where $I_{DS}$ of the target QD is at Coulomb current oscillation peaks.

**FIG. 4.** (a)-(c) Charge stability diagrams of the target QD in three successive voltage ranges of $V_{G2}$. The voltage $V_{G1}$ applied to gate G1 is fixed at $V_{G1}=-0.8$ V. Integer numbers N-1 and N mark the numbers of electrons in the target QD. (d)-(f) Simultaneously measured corresponding transconductance signals $dI_{ds}/dV_{G2}$ of the sensor QD by setting the sensor QD at the declining slope of a Coulomb current peak

as indicated by a green star in Fig. 3(a). (g)-(i) Schematic energy diagrams of the target QD corresponding to the experimental situations as in (a)-(c) and the effective electron occupations in the target QD under different configurations of the energy levels and different polarities of the applied source-drain bias voltage. A hollow circle indicates that the time-averaged electron number in an energy state is zero, a solid black circle indicates that the time-averaged electron number in a state is one, and a gray circle indicates that the time-averaged electron number in a state is a value between zero and one.

**FIG. 5.** (a) Schematic for the charge stability diagram of the target QD around the degenerate N-1 and N electron occupation point. The blue, red, and orange lines mark the critical points where the charge state transitions G(N-1)↔G(N), G(N-1)↔E(N), and E(N-1)↔G(N) take place, respectively. (b) and (c) Schematic energy diagrams of the target QD corresponding to the red and orange lines in (a) in the cases where the tunneling barrier between the target QD and the source electrode is opaquer than the one between the target QD and the drain electrode at high source-drain bias voltages. (d) and (e) Schematic energy diagrams of the target QD corresponding to the red and orange lines in (a) in the cases where the tunneling barriers of the target QD are nearly symmetric at high source-drain bias voltages. Here, hollow, solid black and grey circles have the same meanings as in Fig. 4.

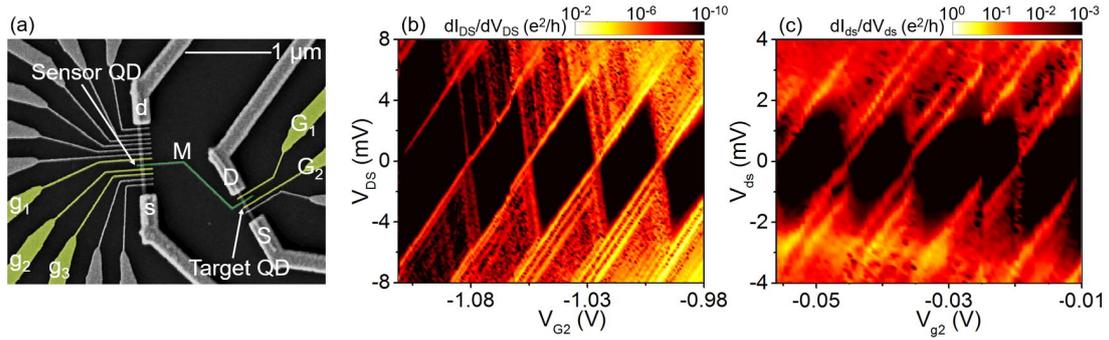

**Figure 1 by Jingwei Mu *et al*.**

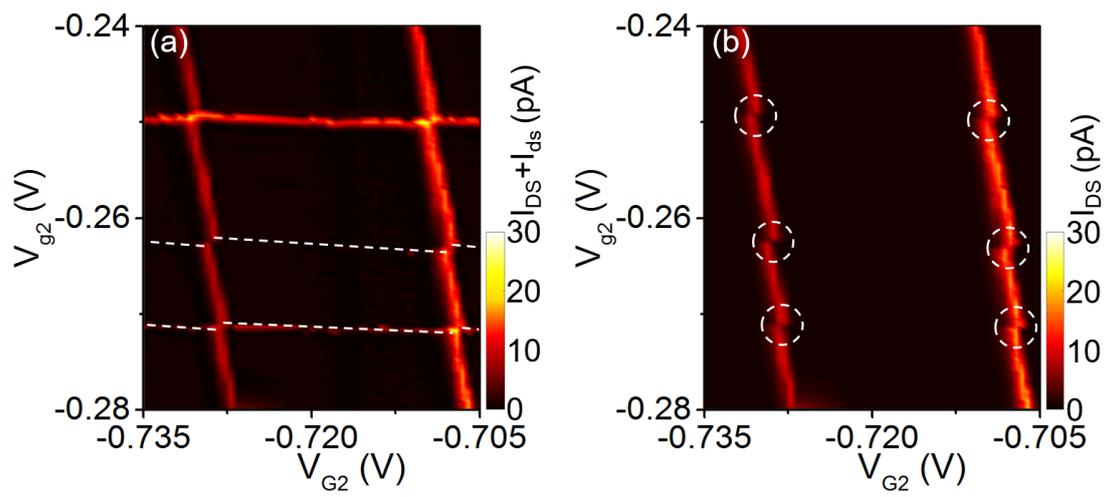

**Figure 2 by Jingwei Mu *et al*.**

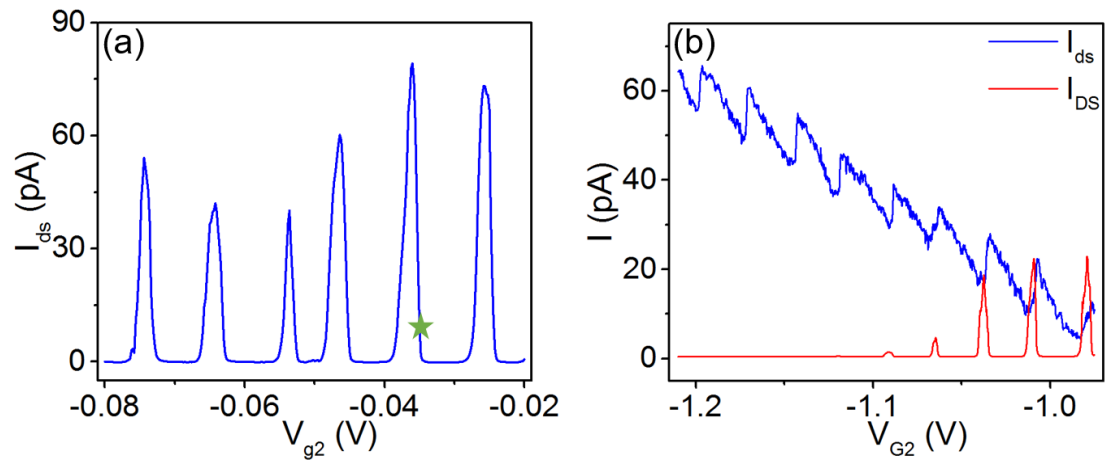

**Figure 3 by Jingwei Mu *et al*.**

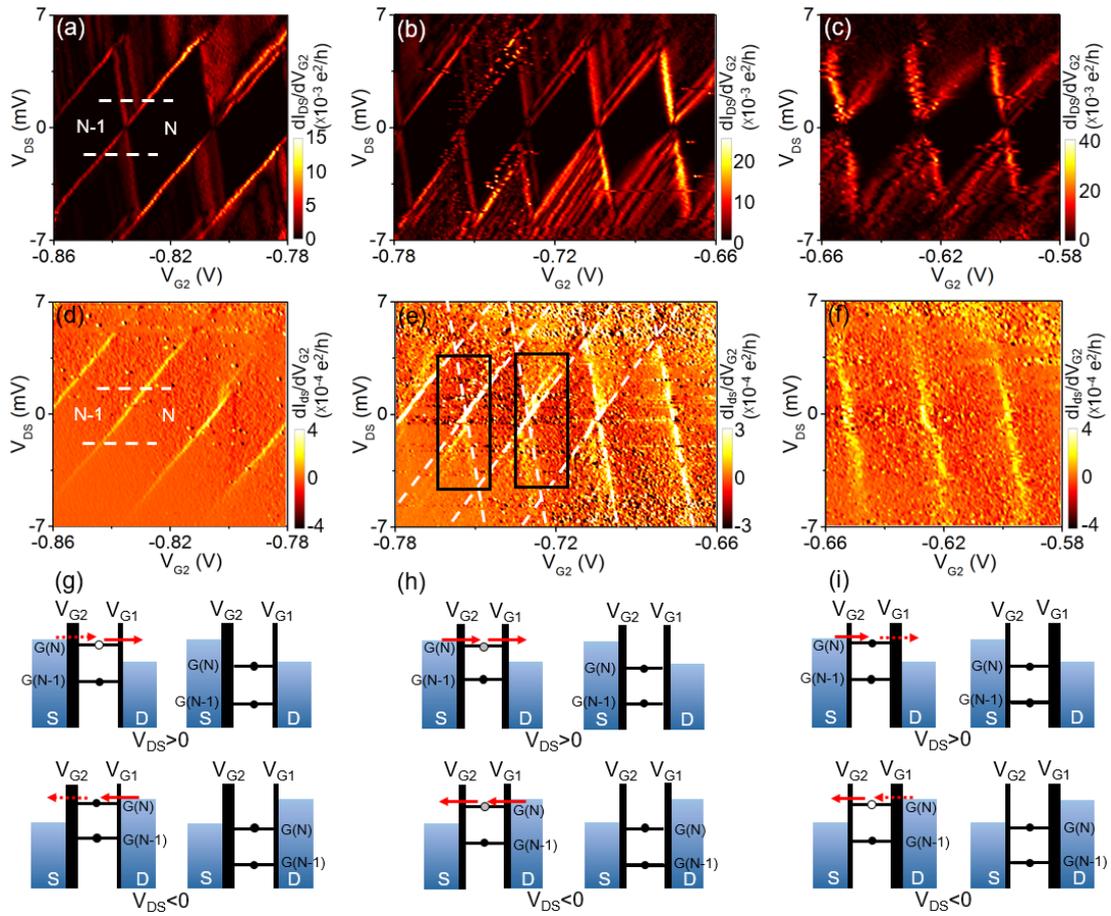

**Figure 4 by Jingwei Mu *et al*.**

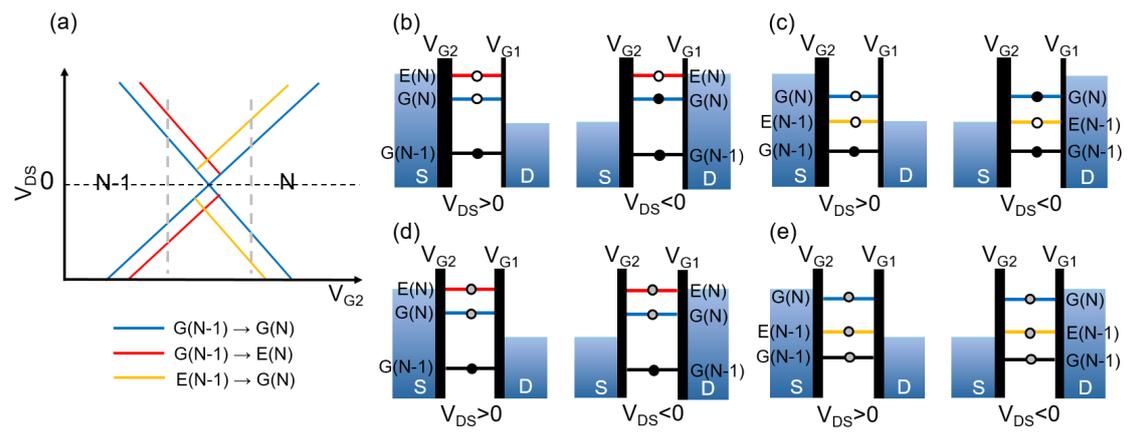

**Figure 5 by Jingwei Mu *et al*.**